\pdfoutput=0
\documentclass{JINST}
\usepackage{textcomp} 

\usepackage{amssymb}
\usepackage{mathrsfs}
\usepackage{graphicx}
\usepackage{dcolumn}
\usepackage{bm}

\title{In situ measurement of the electron drift velocity for upcoming directional Dark Matter detectors}

\author {J.~Billard$^{a,b,c}$, 
F.~Mayet$^a$\thanks{Corresponding author.}, G.~Bosson$^a$, O.~Bourrion$^a$, O.~Guillaudin$^a$, J.~Lamblin$^a$, 
J.~P.~Richer$^a$, Q.~Riffard$^a$, D.~Santos$^a$, F.~J.~Iguaz$^d$, L.~Lebreton$^e$, D.~Maire$^e$.\\
\llap{$^a$}Laboratoire de Physique Subatomique et de Cosmologie,
Universit\'e Joseph Fourier Grenoble 1,
  CNRS/IN2P3, Institut Polytechnique de Grenoble,
  53, rue des Martyrs, Grenoble, France\\
  E-mail: \email{billard@mit.edu}\\
\llap{$^b$} Department of Physics, 
Massachusetts Institute of Technology, Cambridge, MA 02139, USA\\
\llap{$^c$} MIT Kavli Institute for Astrophysics and Space Research, Massachusetts Institute of Technology; Cambridge, MA 02139, USA\\
\llap{$^d$} Universidad de Zaragoza - Departamento de Fisica Teorica, Pedro Cerbuna 12, E-50009 Zaragoza, Spain\\
\llap{$^e$} Laboratoire de M\'etrologie et de Dosim\'etrie des Neutrons, IRSN Cadarache,
13115 Saint-Paul-Lez-Durance, France}

\abstract{Three-dimensional track reconstruction is a key issue for directional Dark Matter detection and it requires a precise knowledge of the
electron drift velocity. Magboltz simulations are known to give a good evaluation of 
this parameter. However, large TPC operated underground
on long time scale may be characterized by an effective electron drift velocity that may differ from the value 
evaluated by simulation. {\it In situ}  measurement
of this key parameter is hence needed as it is a way to avoid bias in the 3D track reconstruction. We present a dedicated method 
for the measurement of the electron drift velocity with the MIMAC detector. It is tested on two gas mixtures :  
$\rm CF_4$ and $\rm CF_4+CHF_3$. 
The latter has been chosen for the MIMAC detector as we expect that adding CHF$_3$ to pure CF$_4$ will lower the electron drift velocity.
This is a key point for directional Dark Matter as the track sampling along the  drift field will be improved while keeping 
almost the same  Fluorine content of the gas mixture.  We show that the drift velocity at $50$~mbar is reduced by a factor of 
about 5 when adding 30\% of $\rm CHF_3$.}

\begin{document}

\section{Introduction}
Directional detection of galactic Dark Matter offers
a unique opportunity to identify Weakly Interacting Massive Particle (WIMP) events as
such~\cite{spergel,billard.exclusion,henderson,morgan1,morgan2,copi1,copi2,copi3,green1,green2,billard.disco,billard.profile,green.disco,billard.ident,Alves:2012ay,Lee:2012pf}.
This new Dark Matter search strategy  requires  the simultaneous measurement of the recoil energy ($E_R$) and the direction of the 
3D track ($\Omega_R$) of  low energy recoils. This can be achieved with low pressure gaseous detectors, in particular Time projection
Chamber (TPC), and  there is a worldwide effort toward the development 
of a large TPC devoted to directional detection \cite{white}. All current 
projects \cite{dmtpc,Lopez:2013ah,drift,d3,mimac,newage} face common challenges amongst which 3D track reconstruction \cite{billard.track} 
is the major one.\\
A key issue for directional Dark Matter detectors is indeed the knowledge of the electron transport 
properties in the gas mixture used as a sensitive medium for the TPC \cite{caldwell}.  
Electron drift velocity is one of the main physical properties used for 3D track reconstruction as 
primary electrons, created along the recoil trajectory, are used to retrieve the recoil track in the TPC. 
In particular, for the MIMAC project \cite{mimac}, the measurement of the third dimension,  along the
electric field, is achieved thanks to a sampling of the primary electron cloud. 
  Large TPC operated underground on long time scale may be characterized by an effective electron drift velocity that may 
  differ from the value evaluated by Magboltz simulation \cite{magboltz}, due to {\it e.g.} impurities, field inhomogeneities, 
  long drift distances. 
{\it In situ}  measurement of this key parameter is hence needed as it is a way 
to avoid bias in the 3D track reconstruction.\\ 
The aim of this paper is to present a dedicated method for an {\it in situ} measurement of the effective 
electron drift velocity within the MIMAC detector 
\cite{mimac}. We emphasize that the goal is not to obtain a precise measurement of the electron drift velocity, to be compared with
simulation for instance, as the experimental set-up is far from being ideal for such a measurement which requires very short drift
distance. On the contrary, we expect a departure from standard values of the electron drift velocity, due to {\it e.g.} large drift
distances, that we aim at measuring with the Dark Matter detector itself. 
 For this purpose, we  use a dedicated experimental set-up, 
including a collimated $\alpha$ source, together with a maximum likelihood method associated to a modelisation of the signal induced 
on the grid, that allows us to estimate the electron drift velocity 
(averaged on the whole detector volume).\\
In most cases, the electron drift measurement is done with a  $\rm N_2$ laser used to generate photo-electrons \cite{schmidt1,schmidt2,Colas.drift}, by measuring the electron 
collection time, between the UV emission time and the electron arrival time on the anode. In order to have a precise measurement,  small drift spaces are 
used \cite{Colas.drift}, $\mathcal{O}$(10) mm, ensuring that the electric field remains homogeneous.  
Nonetheless, the method described in this paper aims at measuring the effective electron drift velocity for a large TPC. 
This is indeed a key point to validate the charge collection within the whole MIMAC 
drift space which is equal to 17.7  cm and 25 cm, respectively for the MIMAC prototype (used hereafter) and the 
forthcoming full-scale MIMAC detector.\\  
The paper is organized as follows. Section \ref{sec:derive:dispositif} presents the detection strategy of the MIMAC experiment together with the dedicated setup for electron drift velocity measurement.
Section \ref{sec:derive:simulation} presents the signal modeling used in the following. 
The new  data analysis strategy, based on a profile likelihood ratio method, is presented in section  
\ref{sec:derive:limite}. Eventually, experimental results obtained with  a pure $\rm CF_4$ gas and 
a $\rm  CF_4 + CHF_3$ gas mixture are presented in section 
\ref{sec:results}.

\section{Experimental setup}
\label{sec:derive:dispositif}

 \begin{figure}[p]
\begin{center}
\includegraphics[scale=0.45,angle=0]{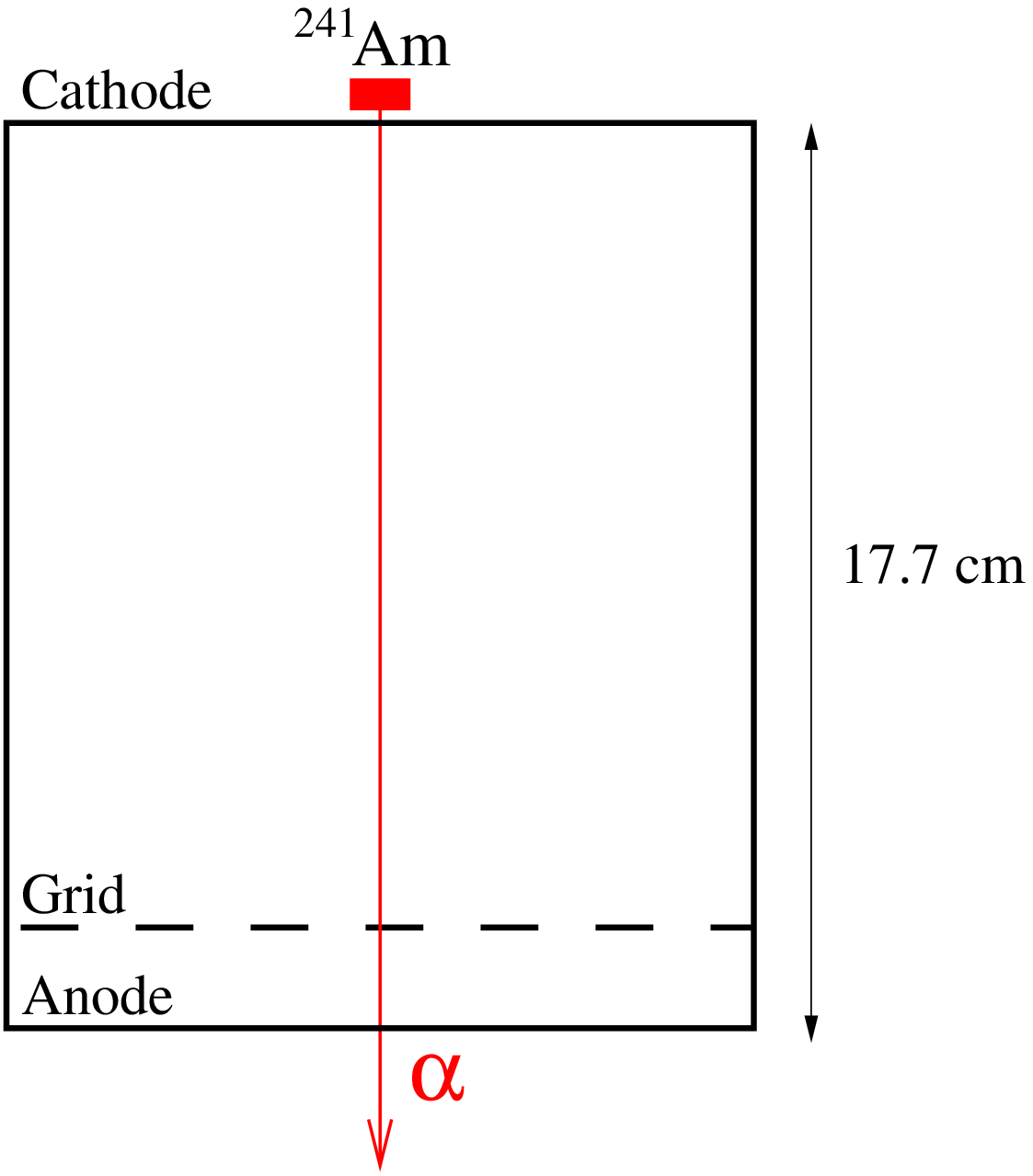}
\caption{\it Sketch representing the experimental setup dedicated to the drift velocity measurement. The collimated alpha source is on the cathode and is facing the anode. To avoid straggling of alpha particles, the latters are going through the cathode thanks to a thin hole. The distance between the cathode and the anode is 17.7 cm and the 
amplification gap is 256 $\mu$m.} 
\label{fig:schema}
\end{center}
\end{figure}

 \begin{figure}[t]
\begin{center}
\includegraphics[scale=0.38,angle=0]{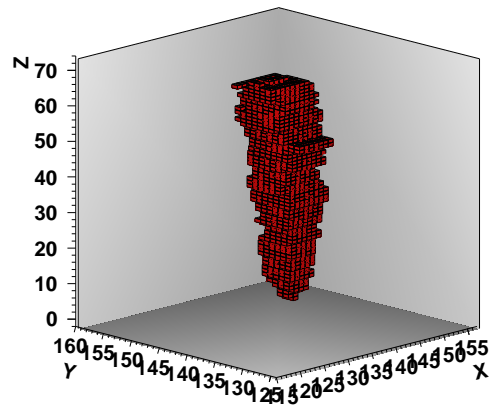}
\caption{\it 3D track of an $\alpha$ particle within the MIMAC detector. The track is crossing the whole drift space, from the cathode to the anode. The X, Y and Z axis are in units of strip number and time sample respectively.} 
\label{fig:onetrack}
\end{center}
\end{figure}

 \begin{figure}[p]
\begin{center}
\includegraphics[scale=0.45,angle=0]{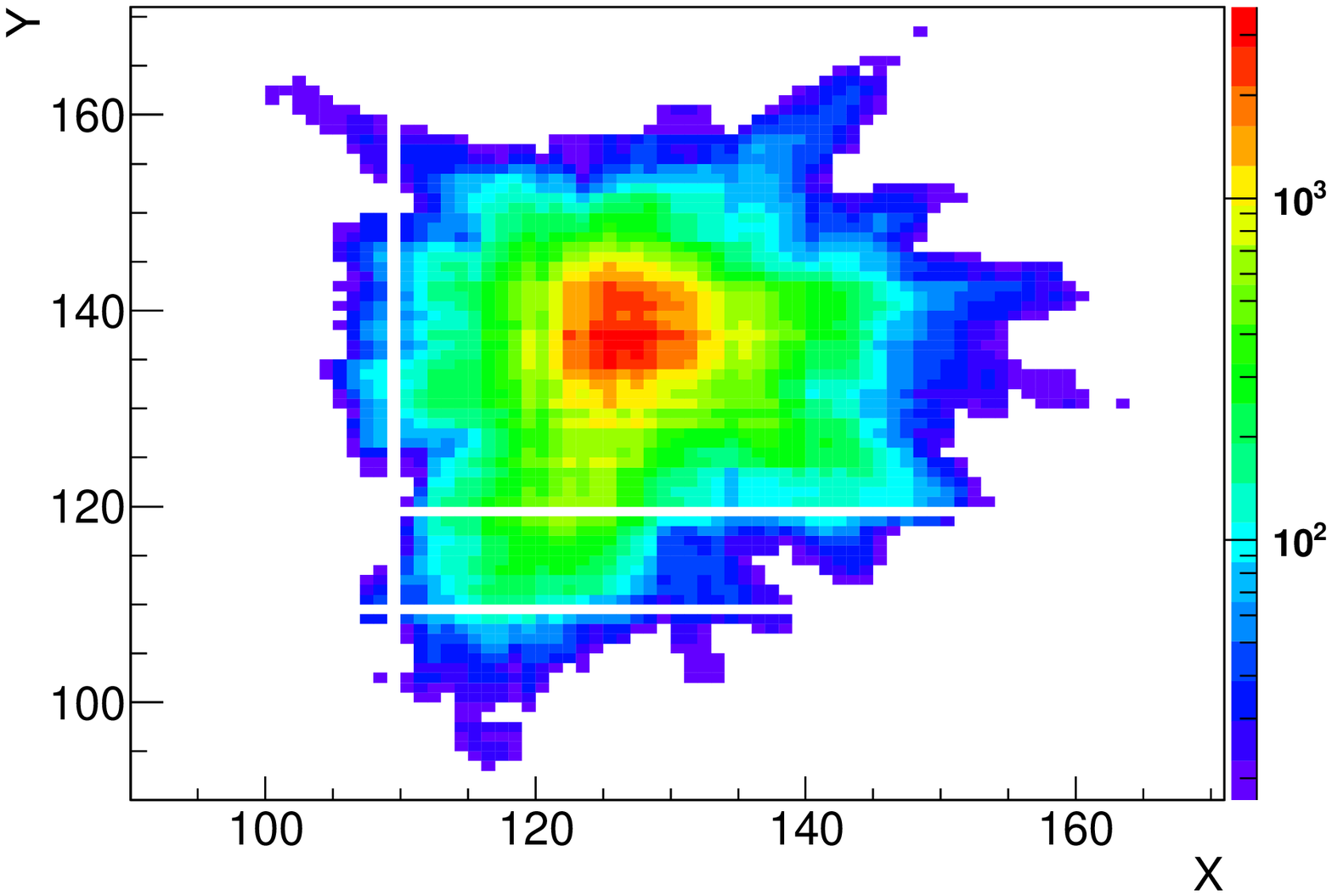}
\includegraphics[scale=0.45,angle=0]{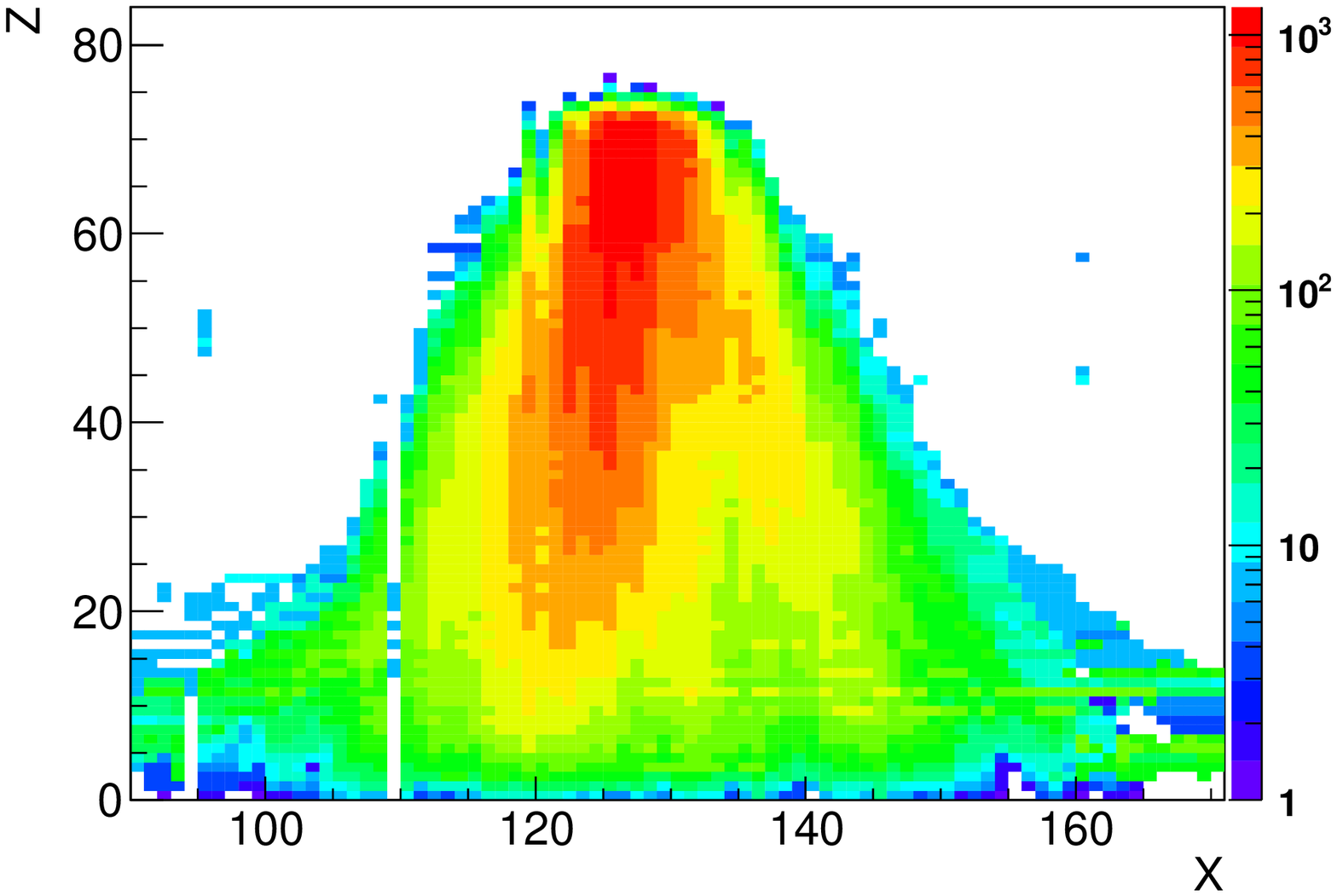}
\caption{\it 2D  projection  of 500 $\alpha$ tracks on the anode plane (upper panel) and on the (X,Z) plane 
(lower panel). The measurement has been done for a pure  CF$_4$ gas  at $50$~mbar, with 
$E_d = 137.9$ V/cm and $E_a = 14.5$ kV/cm. The X, Y and Z axis are in units of strip number and time sample respectively. White lines correspond to dead channels.} 
\label{fig:FULLDriftvelocity}
\end{center}
\end{figure}

\subsection{Measuring low energy nuclear recoils with the MIMAC detector}
The detection strategy of the MIMAC experiment is the following. 
The primary electron-ion pairs produced by a nuclear recoil in the MIMAC chamber are detected by drifting 
the primary electrons to the grid of a bulk Micromegas \cite{giomataris,giomataris2,iguaz} and 
producing the avalanche 
in a very thin gap 256 $\mu$m.  The anode pixelization allows us to get information on the X and Y coordinates.  Indeed, the MIMAC prototype $\mu$TPC is composed of a pixelized anode featuring 2
orthogonal series of 256 strips of pixels (X and Y) [7] and a micromesh grid defining the delimitation
between the amplification (grid to anode) and the drift space (cathode to grid). Each strip of
pixels is monitored by a current preamplifier and the fired pixel coordinate is obtained by using the
coincidence between the X and Y strips (the pixel pitch is 424 $\mu$m).
In order to reconstruct the third dimension of the recoil, the Z coordinate {\it i.e.} along the drift axis, 
 a self-triggered electronics has been developed \cite{Richer:2011pe,Bourrion:2011vk,Bourrion:2011it}. It allows us to perform the anode sampling at a frequency of 50 MHz. Hence, the track is 3D reconstructed, providing the electron drift velocity is known, which is the main interest of this paper.\\
 In parallel to the 3D track measurement, the ionization energy is measured using a charge sensitive preamplifier connected to the grid which is also sampled at a frequency of 50 MHz.

\subsection{Dedicated setup for electron drift velocity measurement}
A dedicated setup has been developed for the electron drift velocity measurement with the MIMAC detector, as 
illustrated in figure~\ref{fig:schema}. We use a collimated $\alpha$ source ($^{241}$Am) producing $\alpha$ particles 
with an average kinetic energy  $E_{\alpha} = 5.478$ MeV.  
With 3 MeV released in a 20 cm $\alpha$ track, it corresponds to an  average $dE/dx$  close the one of a 
nuclear recoil leaving about 100 keV in 5 mm. Moreover, as the ion drift velocity in the avalanche gap is close to the one of electrons in
the drift gap, we   do not expect an effect from space charge. The source is positioned on the cathode and facing the anode. 
As the energy loss  is about 3 MeV in the TPC ($50$~mbar of CF$_4$), the $\alpha$ particle is expected to cross the whole drift space (17.7 cm)
as well as the 256 $\mu$m amplification space. As the $\alpha$ particle velocity is much greater than the electron drift velocity by about 2 orders of magnitude, the $\alpha$ arrival
time on the anode is taken as the starting time of primary electrons at the cathode.\\  
Figure~\ref{fig:onetrack} presents a typical 3D track for an $\alpha$ particle that is used for the electron drift velocity measurement. 
The track is crossing the whole drift space, from the cathode to the anode. 
Figure~\ref{fig:FULLDriftvelocity}  presents, for $\sim 500$ $\alpha$ particles, the projection on the (X,Y) anode plane  (upper panel) and on the (X,Z) plane (lower panel). 
It can be seen that the $\alpha$ source is a pencil point-like one, with a 5$^{\circ}$ opening angle, which ensures that we are sensitive to the primary electrons created at the cathode. The latter is a necessary condition to validate our measurement strategy based on the drift time of primary electrons created along the track in the detector, from the anode to the cathode.\\
 
As discussed in the next section, it is compulsory to be able to measure the time dependent charge collection profile to recover a non biased estimate of the drift velocity. The rise time of the charge sensitive preamplifier (when charges are injected on the grid), is about 400 ns. This is much lower than the expected collection time of the primary electrons coming from the alpha tracks for the different gases and drift field considered hereafter, which are estimated using 
Magboltz to be between 1.5 $\mu$s and 17 $\mu$s for a pure CF$_4$ gas and a CF$_4$ + CHF$_3$ gas mixture respectively, see sec.~\ref{sec:derive:resultat}. Hence, we should be able to measure an accurate charge collection profile for each alpha track observed.\\
As the alpha particle will deposit an ionisation energy around 3 MeV in the detector, unlike a low energy nuclear recoil measurement, 
we do not need a high gain from the avalanche. So we have slightly modified the energy range of the MIMAC detector in 
lowering  the usual amplification field by a factor of 1.5, hence to a value of 14.5 kV/cm.

\section{Signal modeling}
\label{sec:derive:simulation}
The analysis method proposed in section \ref{sec:derive:limite} requires a complete modeling of the signal, from the current induced on the grid to the measured signal  $V(t)$.\\ 
We first describe the modeling of the current induced on the grid by an $\alpha$ particle crossing the drift space in which electron-ion pairs are created.
The charge induced on the grid will therefore depend on the  time evolution of the electron collection.
In order to reproduce experimental data, we have used the ionisation energy loss  $dE/dx$ simulated  with Geant 4 \cite{geant4}. For a given electron drift 
velocity $v_d$, it corresponds to a time projection, $dE/dt$, of the collection of primary electrons on the anode. As a matter of fact, 
the induced current on the grid is mainly caused by the motion of ions created from the avalanche that are being collected on the grid. The induced current will last for a time  $\Delta t_{\rm ion} =  \epsilon / v_{d_{\rm ion}}$, where $v_{d_{\rm ion}}$ is the drift velocity of ions in the amplification space and 
$\epsilon$ is the Micromegas amplification width  (256 $\mu$m) \cite{giomataris,giomataris2,iguaz}. 
Each primary electron  induced current is associated with an ion-induced  current $Q_{\rm ion}(t)$,  that can be well approximated by a gate function of width $\Delta t_{\rm ion}$, that must be convolved with the anode 
charge collection $dE/dt$. It results in a lengthening  of the signal. Note that the drift velocity of ions is about 2 to 3 orders of magnitude lower than for electrons \cite{sauli}. The latters are expected, from Magboltz simulations, to be about 2 orders of magnitude faster in the amplification space than in the drift space. Hence, one can roughly  expect that the ion drift velocities in the amplification space should be about $\mathcal{O}(10)$ $\mu$m/ns for the considered experimental conditions in sec.~\ref{sec:results}.\\ 

Eventually, the electron longitudinal diffusion must be accounted for as it may significantly lengthen the charge collection profile. 
To take this effect into account, the signal is convolved with a gaussian distribution $g_{\rm diff}(t)$, 
having  a standard deviation 
$\sigma_l(t) = D_l\sqrt{v_d\times t}$, where $D_l$ is the longitudinal diffusion coefficient.  
The current induced on the grid is thus given by: 
\begin{equation}
I_{\rm ind}(t) \propto \int\int \frac{dE}{dt}(t - \xi) \times Q_{\rm ion}(\xi - \tau) \times g_{\rm diff}(\tau)\ d\tau d\xi 
\end{equation}
According to Magboltz simulations, for the following experimental conditions considered in sec.~\ref{sec:results},  the longitudinal diffusion coefficient is expected to lie within the range of 300 to 600 $\mu$m/$\sqrt{\rm cm}$.\\

\begin{figure}[t]
\begin{center}
\includegraphics[scale=0.45,angle=0]{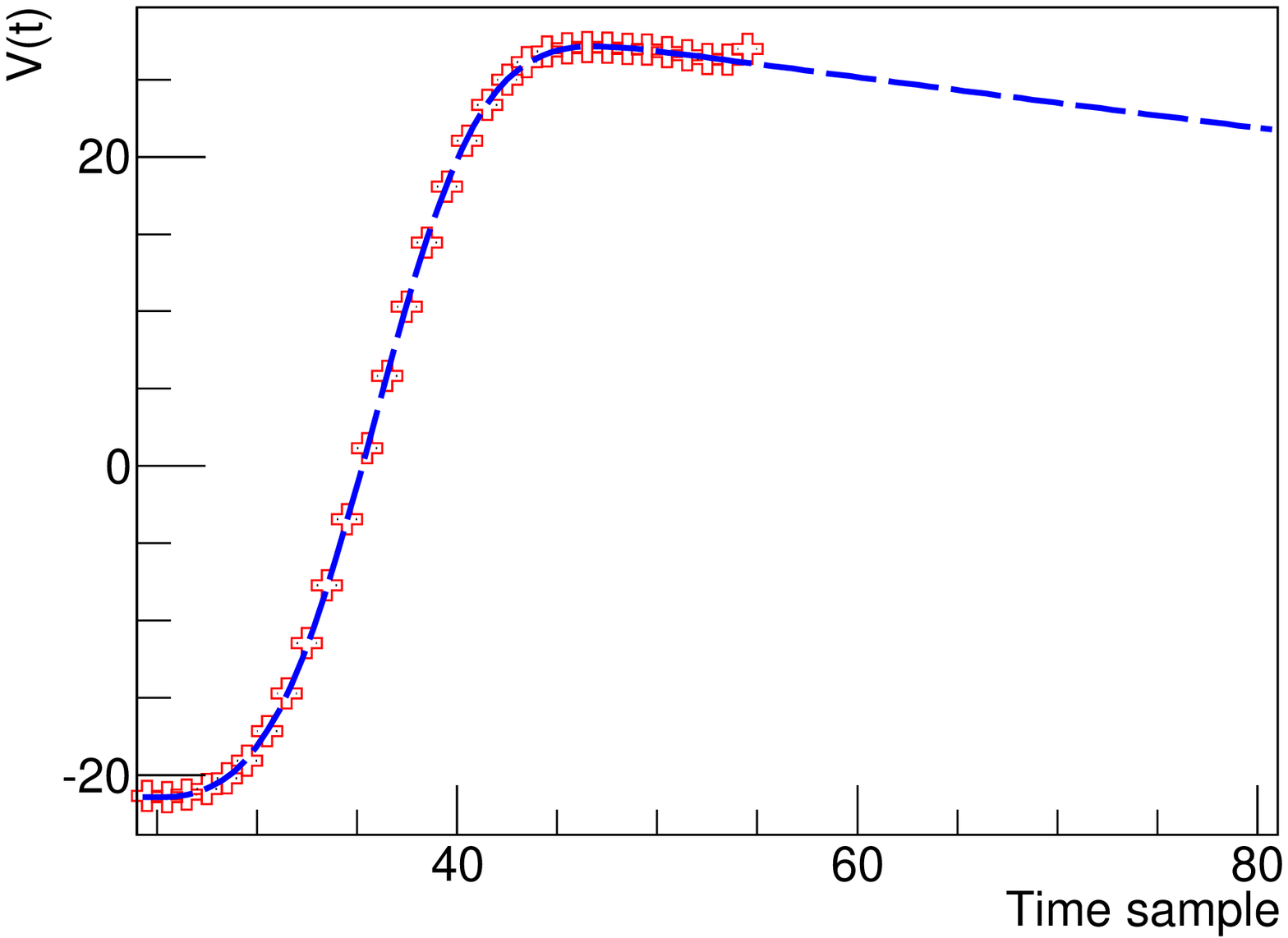}
\caption{\it Transfer function of the charge sensitive preamplifier: measurement (red crosses), analytical expression (blue dashed line).} 
\label{fig:SimuSignal}
\includegraphics[scale=0.45,angle=0]{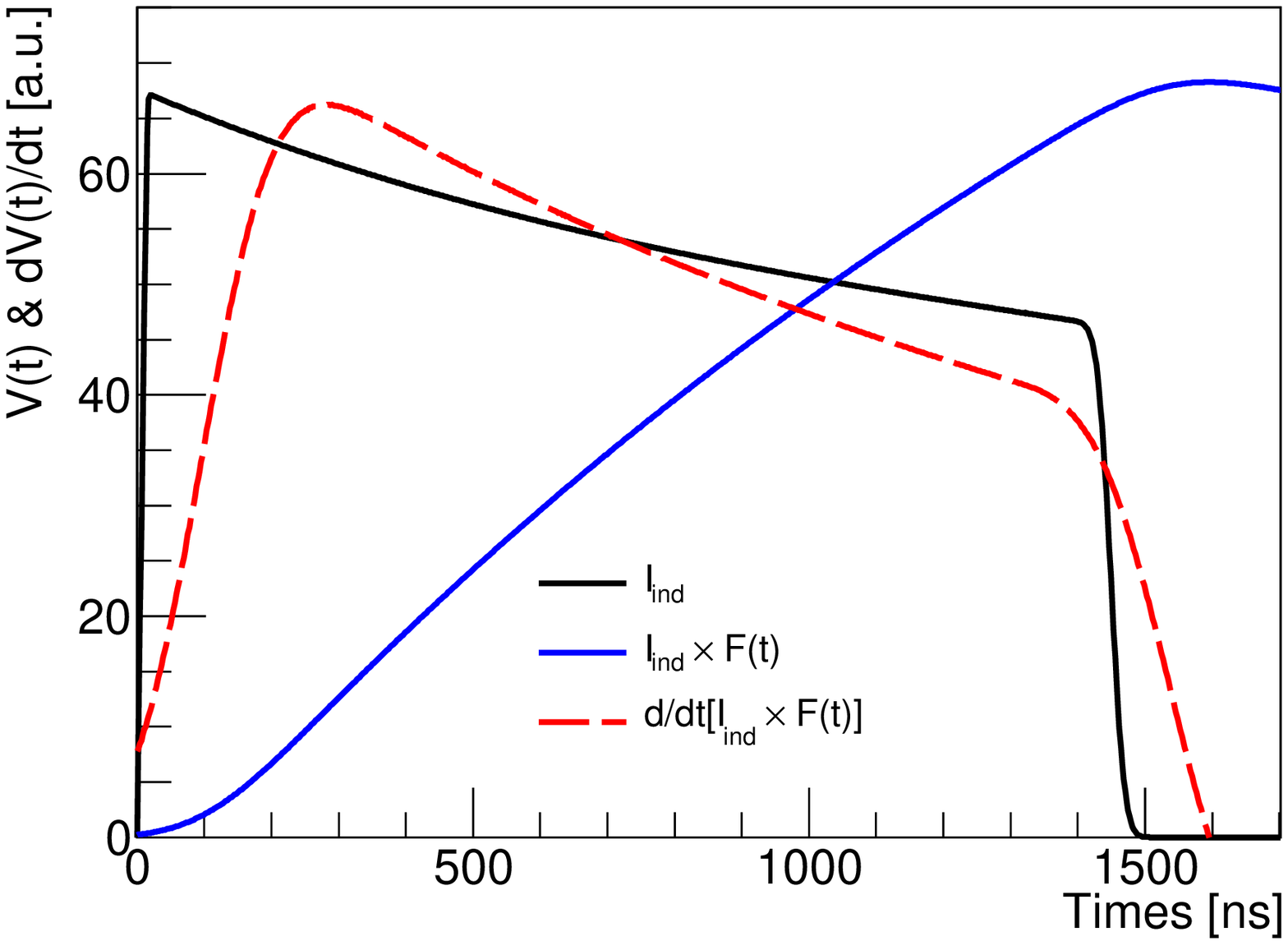}
\caption{\it Theoretical signal $V_{\rm th}(t)$ (blue curve) obtained from the convolution of the current induced on the 
grid $I_{\rm ind}(t)$ (blaks curve) and the transfer function of the charge preamplifier $F(t)$. 
The dashed red curve is the time derivative  $V_{\rm th}^\prime(t)$  of the signal $V_{\rm th}(t)$.} 
\label{fig:Convol}
\end{center}
\end{figure}

The next step is to convolve our induced current model with the charge sensitive preamplifier transfer function\footnote{More precisely it corresponds to the transfer function of the charge sensitive preamplifier and  the  electronic readout.} in order to get an accurate simulation of the output voltage $V(t)$ for each alpha track. 
The measurement of the time response of the preamplifier is done by inducing current pulses directly on the grid with a time width far smaller than the rise time of the preamplifier, which is about 400 ns. The measured transfer function $F(t)$ of the charge sensitive preamplifier in the time domain sampled at a frequency of 50 MHz is shown on figure~\ref{fig:SimuSignal} as the red crosses. In order to convolve the simulated induced current with the preamplifier transfer function, we used an analytical expression of $F(t)$ that is also shown on figure~\ref{fig:SimuSignal} as the blue dashed curve. 
One can see that our analytical model of the charge sensitive preamplifier transfer function is well approximated by our analytical expression.\\

The resulting theoretical signal $V_{\rm th}(t)$ is thus given by 
\begin{equation}
V_{\rm th}(t) \propto \int\int\int \frac{dE}{dt}(t - \xi) \times Q_{\rm ion}(\xi - \tau) \times g_{\rm diff}(\tau -T) \times F(T)\ dT d\tau d\xi
\end{equation}
Figure~\ref{fig:Convol} presents the theoretical signal $V_{\rm th}(t)$ (blue curve) obtained from the convolution of the induced current on the grid $I_{\rm ind}(t)$ (black curve) and the 
transfer function of the charge preamplifier $F(t)$  (see fig.~\ref{fig:SimuSignal}). The time derivate  $V_{\rm th}'(t)$ of the signal $V_{\rm th}(t)$ is also presented (dashed curve). 
For this example, this figure has been plotted considering the following parameter values: 
$D_l = 440 \ \mu {\rm m/\sqrt{cm}}$, $v_{d_{\rm ion}} \sim 8 \ \mu {\rm m/ns}$ and 
$v_d = 122 \ \mu {\rm m/ns}$ (see sec.~\ref{sec:derive:vraisemblance:illustration}). As one can see from figure~\ref{fig:SimuSignal}, the addition of the ion drift velocity, the electron diffusion and the electronic readout introduces a significant lengthening of the signal. From the induced current simulation, one can see that the last electron arrives 1400 ns after the first one while the maximum of the output voltage is at 1600 ns. Note that 
the delay cannot be simply subtracted as it is dependent on the setup configuration. \\
The theoretical signal  depends on $v_d,v_{\rm ion},D_l$ and reads hereafter as 
\begin{equation}
V_{\rm th}(t;v_d,v_{\rm ion},D_l)
\end{equation}
This function, that can be evaluated for any value of the set of parameters ($v_d$, $v_{d_{\rm ion}}$ and $D_l$), is the adjusting model used in the following likelihood function dedicated to the evaluation of the electron drift velocity.

\section{Data analysis strategy}
\label{sec:derive:limite}
Straightforward data analysis strategies would not allow us to estimate the electron drift velocity without bias. After discussing the bias
expected with simple data analysis, we present a data analysis 
strategy based on a profile likelihood method that avoid bias  due to electron diffusion, ion collection time and electronic readouts.
sign
\subsection{Extracting an electron drift velocity from data}
Several data analysis strategies may be used to retrieve the electron drift velocity from the data obtained with this experimental setup.
A straightforward data analysis strategy consists in using the electron drift time   corresponding to a drift length $d = $17.7 cm.  The electron drift velocity $v_d$ is then simply estimated as 
\begin{equation}
v_d = \frac{d}{\Delta t_e}
\end{equation}
with $\Delta t_e$ being the time difference between the $\alpha$ particle arrival time (on the anode) and the last primary electron (generated at the cathode level).\\
This time difference may be estimated with the charge preamplifier, connected to the grid, by measuring the time between the maximum and the minimum.
However, this method allows only for a rough estimate of the electron drift velocity. In fact and   as previously discussed, mostly due to  the readout time constant we expect significant lenghtening of the output signal, leading to an
under-estimation of $v_d$.\\

The second data analysis strategy consists in using the information contained in the 3D track of the $\alpha$ particle.
Indeed, as the strips of the pixelized anode are linked to current preamplifiers, their electronic signal is not delayed (rise time of a few nanosecond), thus allowing for a better estimation of the electron time collection. 
The electron drift velocity $v_d$ may then be  
estimated as 
\begin{equation}
v_d = \frac{d}{\Delta t_c}
\end{equation}
where $\Delta t_c$ is the time difference between the first and the last spatial coincidence. While being almost not delayed by the electronic readout, this estimation is expected to depend  
heavily on the amplification electric field (the gain).  Indeed, the probability to have a 
spatial coincidence depends on the number of electrons contained in a given time sample and hence on the amplification gain.

Eventually, we have checked that none of these methods allows us to estimate the electron drift velocity without bias and in a robust way. Hence, we propose an 
analysis strategy  based on a likelihood method, allowing to avoid bias due to
electron diffusion, ion collection time and electronic readouts.\\

\subsection{A likelihood-based data analysis}
\label{sec:derive:vraisemblance}
\begin{figure}[t]
\begin{center}
\includegraphics[scale=0.5,angle=0]{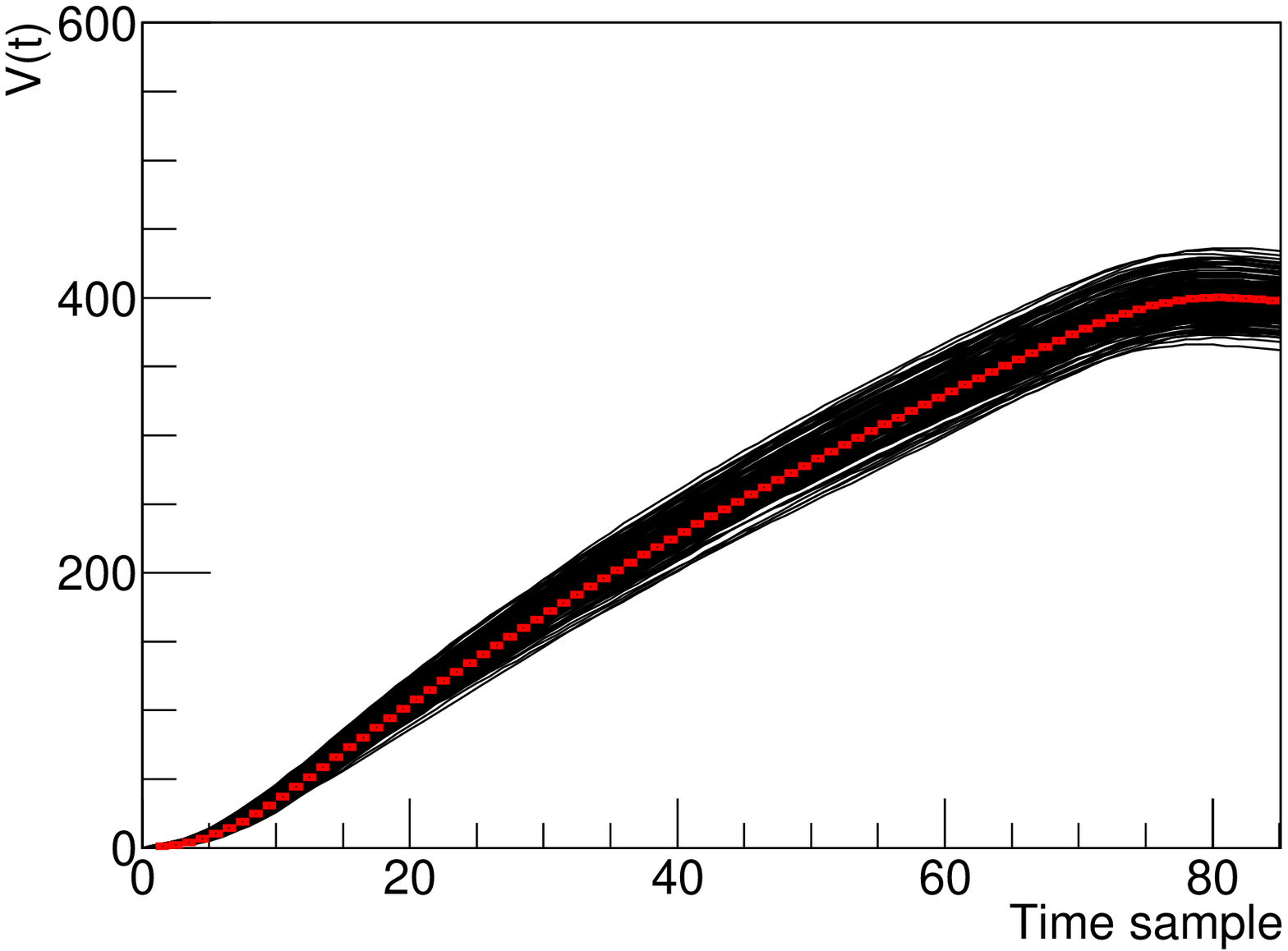}
\caption{\it 500  $V(t)$ profiles as 
well as the mean profile $\bar{V}(t)$ (red). The measurement has been done for a pure  CF$_4$ gas   at $50$~mbar, with 
$E_d = 137.9$~V/cm and $E_a = 14.5$~kV/cm.} 
\label{fig:Integ}
\end{center}
\end{figure}

\begin{figure}[t]
\begin{center}
\includegraphics[scale=0.45,angle=0]{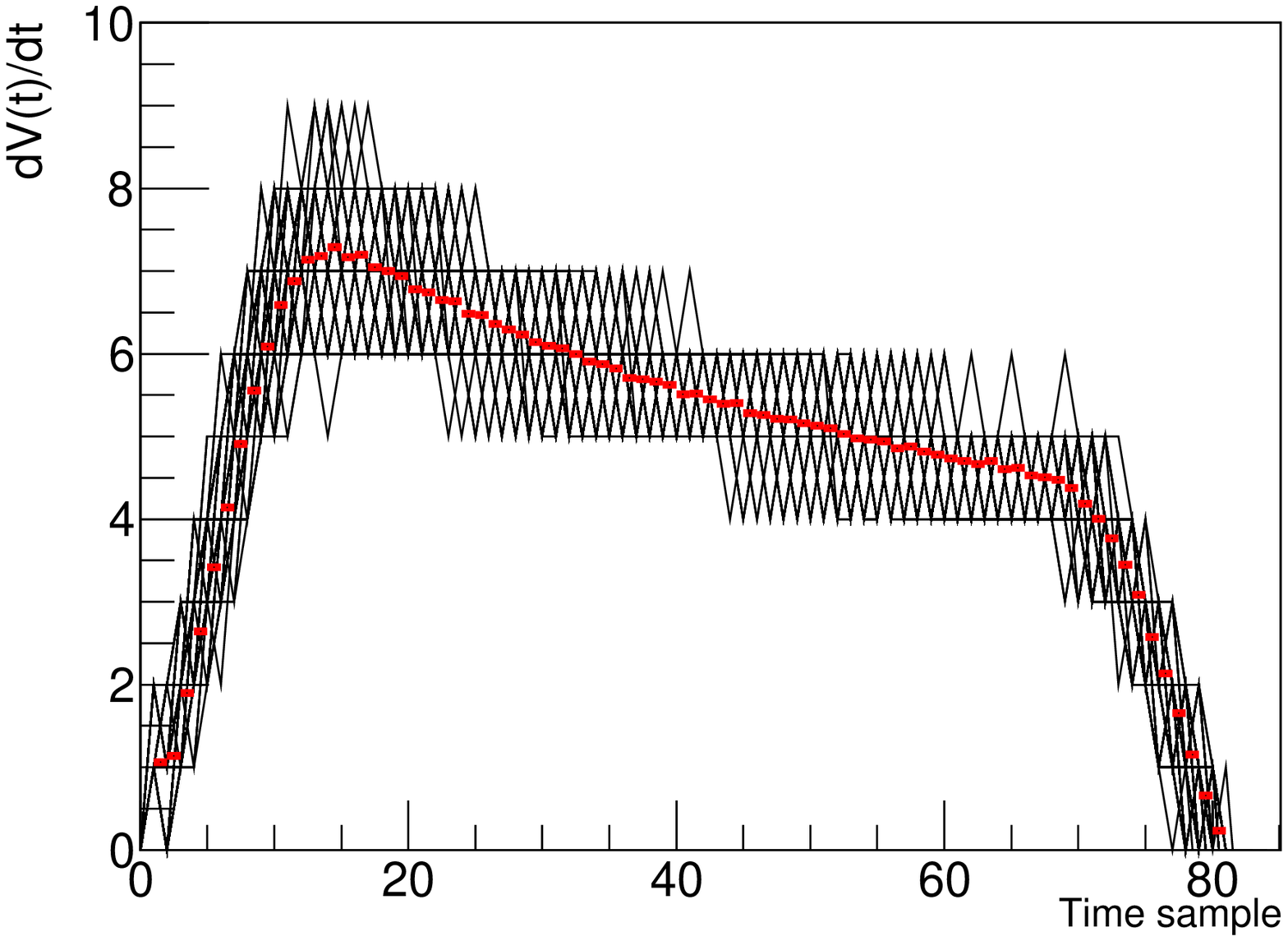}
\includegraphics[scale=0.45,angle=0]{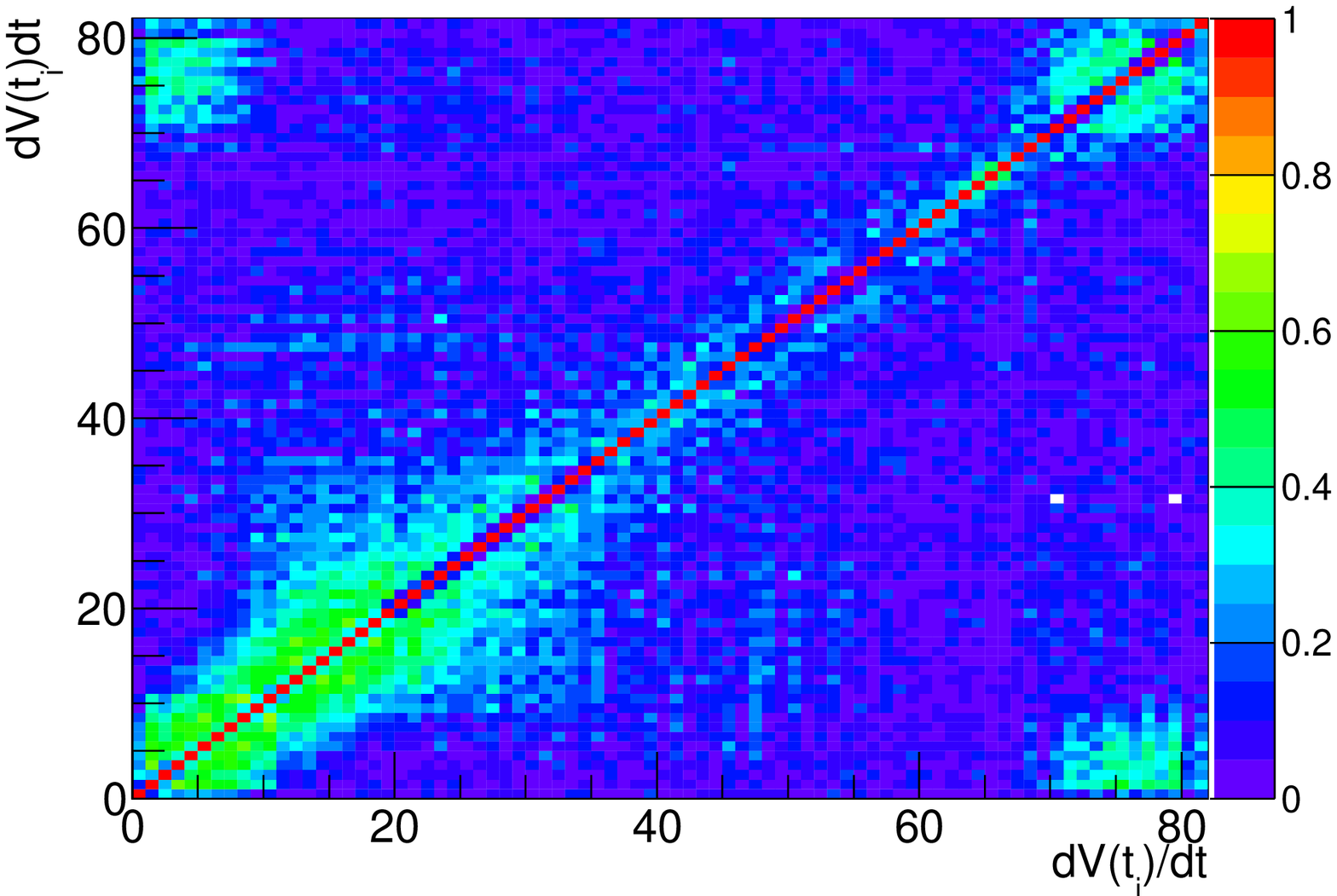}
\caption{\it  Upper panel:  500  $V^\prime(t)$ profiles as 
well as the mean profile $\bar{V^\prime}(t)$ (red). Lower panel:  the correlation matrix  
$\rho[V^\prime(t_i),V(t_j)]$. The measurement has been done for a pure  CF$_4$ gas  at $50$~mbar, with 
$E_d = 137.9$ V/cm and $E_a = 14.5$ kV/cm.}
\label{fig:Deriv}
\end{center}
\end{figure}

The likelihood data analysis method is based on a comparison between experimental data and simulation obtained from the previously discussed signal modeling 
(sec.~\ref{sec:derive:simulation}). 
For each setup configuration, we measure   $\sim 500$ $\alpha$ tracks, in order to minimize statistical uncertainties in
the estimation of $v_d$. Using the $\sim 500$ profiles of $V(t)$, we evaluate a mean profile $\bar{V}(t)$ that is being adjusted by the  
 signal model $V_{\rm th}(t;v_d,v_{\rm ion},D_l)$. Figure~\ref{fig:Integ} presents the 500  $V(t)$ profiles (in black) as 
well as the mean profile $\bar{V}(t)$ (red). 
However, evaluating the likelihood function as a product of likelihoods requires that each $V(t_i)$ are independent of 
each other. This is obviously not the case as the  $V(t)$ signal corresponds to an integration of the induced current. Hence the
$V(t_{i+1})$ value  is directly related to the  $V(t_{i})$ one. In particular, we have checked that the correlation matrix, 
$\rho[V(t_i),V(t_j)]$, is highly non diagonal. To avoid a diagonalization of the covariance matrix,  the time derivate of
the signal, $V^\prime(t)$, is used instead. The upper panel of figure~\ref{fig:Deriv} presents the 500  $V^\prime(t)$ profiles as 
well as the mean profile $\bar{V^\prime}(t)$ (red) while the lower one presents the correlation matrix, 
$\rho[V^\prime(t_i),V^\prime(t_j)]$. It can be noticed that the  $V^\prime(t_i)$ are only weakly correlated with each other. 

Having demonstrated that  $V^\prime(t_i)$  are weakly correlated, 
the likelihood function can be written as the product of the likelihoods associated with 
each value of  $\bar{V^\prime}(t_i)$. It reads as 
\begin{equation}
\mathscr{L}(v_d,v_{\rm ion},D_l,\delta t, A) = \exp\left(-\frac{1}{2}\sum_{i=1}^{N_t}\left[\frac{A\times V_{\rm th}'(t_i-\delta t;v_d,v_{\rm ion},D_l) - \bar{V}'(t_i)}
{\sigma_{\bar{V}'(t_i)}}   \right]^2\right)
\end{equation}
where $\delta t$ et $A$ are adjusting parameters, to enable a time and amplitude shift  
between the data and the
adjusting model.  $\bar{V^\prime}(t)$ is the mean profile value, $\sigma_{\bar{V^\prime}'(t)}$ its statistical standard deviation and $N_t$ is the number of time samples.
It is worth noticing that the four nuisance parameters are associated with flat and non informative prior distributions. 

\section{Experimental results}
\label{sec:results}
This method for {\it in situ} electron drift velocity measurement   has been applied  to two gas mixtures 
that might be used for directional Dark Matter detection: pure $\rm CF_4$ and  $\rm  CF_4 + CHF_3$. In the following, we will always consider a gas pressure of $50$~mbar.
 
\subsection{Illustration of the method}
\label{sec:derive:vraisemblance:illustration}
\begin{figure}[p]
\begin{center}
\includegraphics[scale=0.45,angle=0]{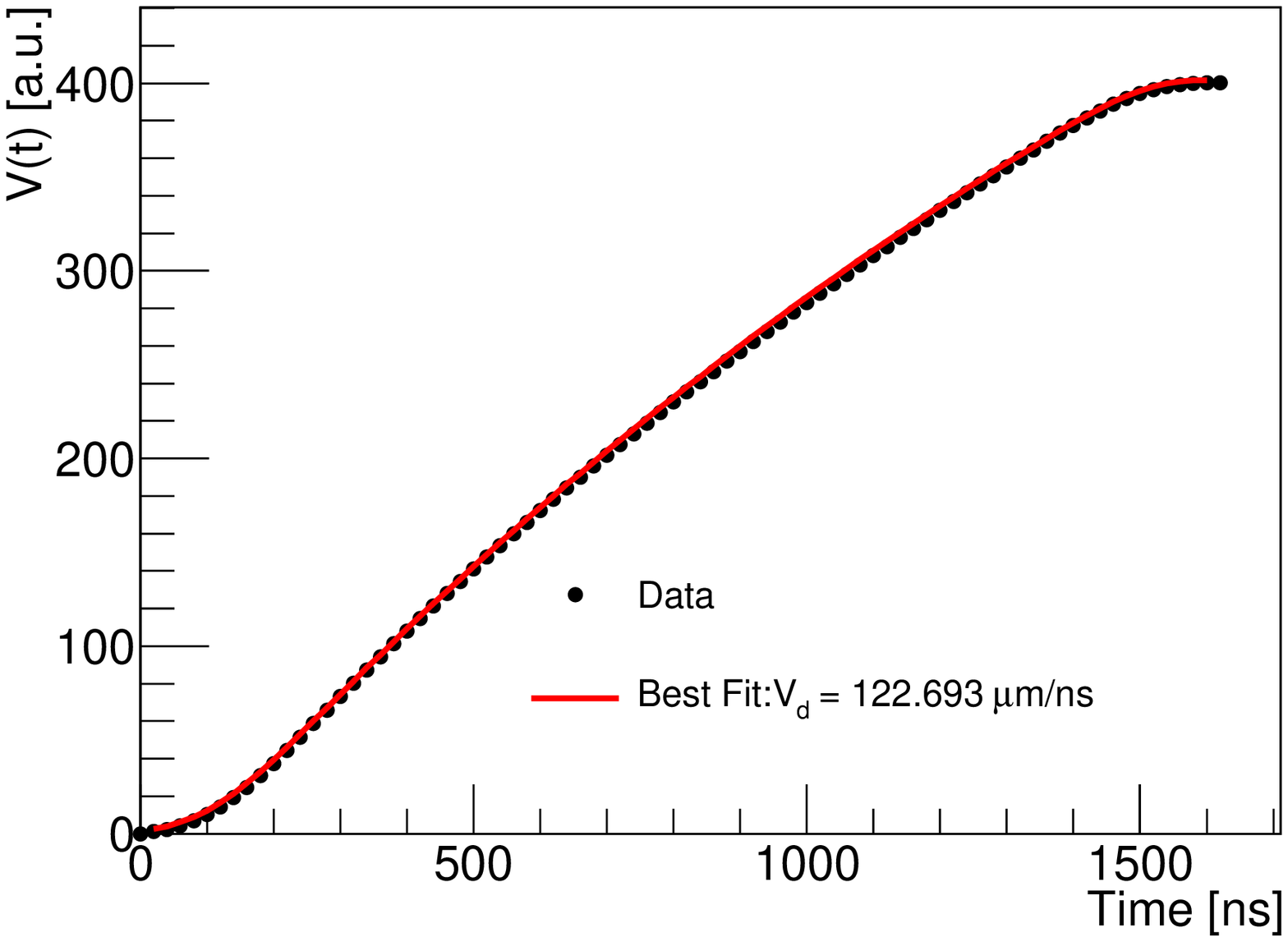}
\includegraphics[scale=0.45,angle=0]{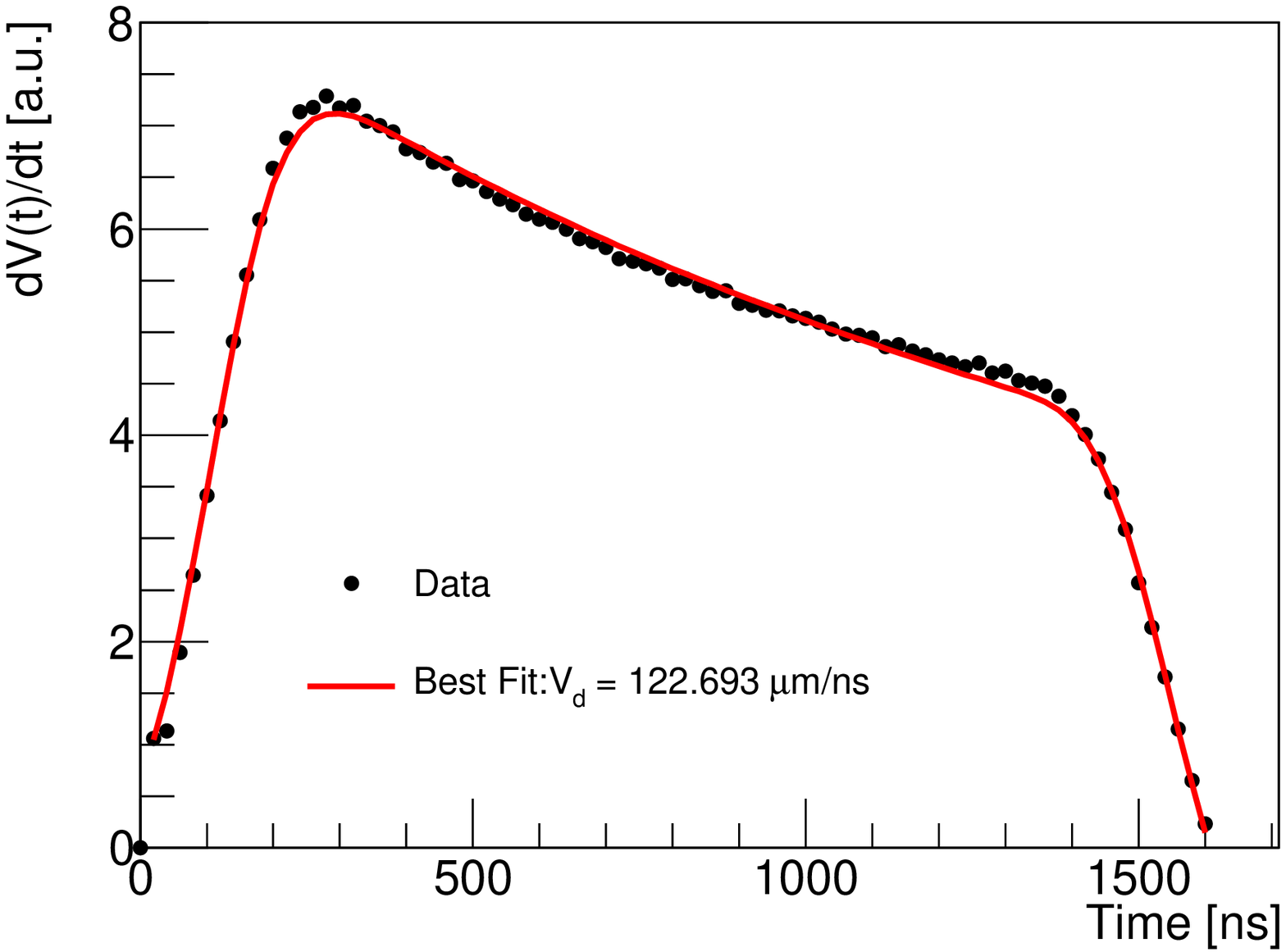}
\caption{\it Mean profile (black point) and  best fit (red curve) for the signal ${V}(t)$ (upper panel) and its time derivate  
$V^\prime(t)$ (lower panel). The measurement has been done for a pure  CF$_4$ gas  at $50$~mbar, with 
$E_d = 137.9$ V/cm and $E_a = 14.5$ kV/cm.} 
\label{fig:BestFit}
\end{center}
\end{figure}

To examplify this experimental method, we present the full result for the same experimental setup as for fig.~\ref{fig:Integ} and ~\ref{fig:Deriv}, namely 
pure $\rm CF_4$ at $50$~mbar and a drift electric field of $\rm E_d = $ 137.29 V/cm. The mean profiles ($\bar{V}(t)$ and $\bar{V^\prime}(t)$) are presented on 
figure~\ref{fig:BestFit} (black points). The result of the maximization of the likelihood function  
$\mathscr{L}(v_d,v_{d_{\rm ion}},D_l,\delta t, A)$ in this case is presented as a red curve. 
As outlined in section 
\ref{sec:derive:vraisemblance}, the likelihood maximization is performed on the  $\bar{V^\prime}(t)$ profile. Hence, the comparison 
with the $\bar{V}(t)$ profile is only a consistency check. The adjustment is excellent, in
particular in the regions of interest for the estimation of the electron  drift  
velocity, {\it i.e.} rising and falling part of the  mean profile $\bar{V^\prime}(t)$. This emphasizes the fact that no space charge effect
is observed. Small differences between the  fit 
and the data    in the central region ($300 \ {\rm ns} \leq t \leq 1200 \ {\rm ns}$) are due to a lack of accuracy 
in the estimation of the decreasing part of the preamplifier transfer function.\\

\begin{figure}[t]
\begin{center}
\includegraphics[scale=0.45,angle=0]{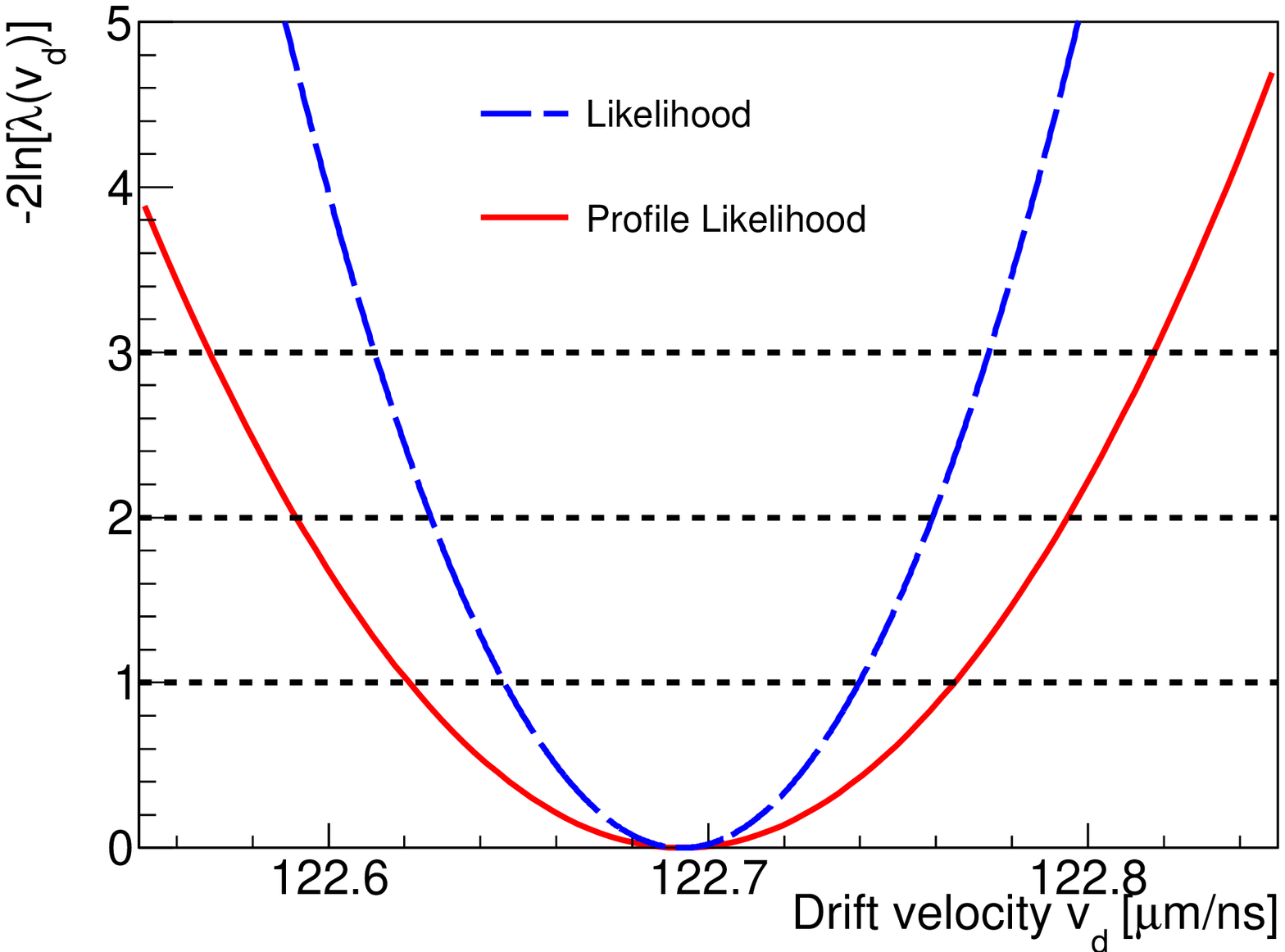}
\caption{\it Likelihood ratio ($-2\ln[\lambda({v}_d \pm \sigma^{\pm}_{v_d})]$) as a function of the 
electron  drift  velocity $v_d$ for a   likelihood function (blue) and a  profile likelihood functon (red). 
The measurement has been done for a pure  CF$_4$ gas  at $50$~mbar, with 
$E_d = 137.9$ V/cm and $E_a = 14.5$ kV/cm.} 
\label{fig:profileVelocity}
\end{center}
\end{figure} 

The use of a likelihood method allows us to estimate the uncertainty on the electron  drift  velocity thanks to the use 
of the standard profile likelihood ratio test statistic. Indeed, we evaluate the likelihood ratio 
$\lambda(v_d)$ as a function of the parameter $v_d$, as 
\begin{equation}
\lambda(v_d) = \frac{\mathscr{L}(v_d,\hat{\hat{v}}_{d_{\rm ion}},\hat{\hat{D}}_l,\hat{\hat{\delta t}}, \hat{\hat{A}})}{\mathscr{L}(\hat{v}_d,\hat{v}_{d_{\rm ion}},\hat{D}_l,
\hat{\delta t}, \hat{A})}
\end{equation}
where the double hat notation corresponds to the maximum of the conditional likelihood, when one of the parameters are taken at fixed
value. Then, the uncertainty at the 68\% confidence level on $v_d$ is obtained by solving 
\begin{equation}
-2\ln[\lambda(\hat{v}_d \pm \sigma^{\pm}_{v_d})] = 1
\end{equation}
where  $\sigma^{\pm}$ is the upper and lower asymmetric error bars. \\

Figure \ref{fig:profileVelocity} presents the likelihood ratio ($-2\ln[\lambda({v}_d \pm \sigma^{\pm}_{v_d})]$) as a function of the 
electron  drift  velocity $v_d$ with (red) and without (blue) profiling over the other fitting parameters. For the latter case, all
parameters, except  $v_d$ , are taken at the  value estimated by the maximization of the likelihood function. As expected, the uncertainty obtained with the profiled likelihood function is larger 
(by $\rm \sim 40 \ \%$) as the uncertainties on all the other parameters are taken into account.
On this example, we found:
\begin{equation}
v_d = 122.7 \pm 0.14 \ \mu {\rm m/ns} \ \ \ (68\%  \ {\rm C.L.})
\end{equation}
This result leads to a measurement of the electron drift velocity with a precision of  $\sim 0.1 \ \%$. However, one caveat of our likelihood method is that it does not include systematics from the experimental setup, such as the length of the chamber
and the homogeneity of the electric field along the drift space. These systematics can be included afterwards and conservatively evaluated to be around 1\%.

\subsection{Results from the drift velocity measurement with MIMAC}
\label{sec:derive:resultat}

\begin{figure}[p]
\begin{center}
\includegraphics[scale=0.6,angle=0]{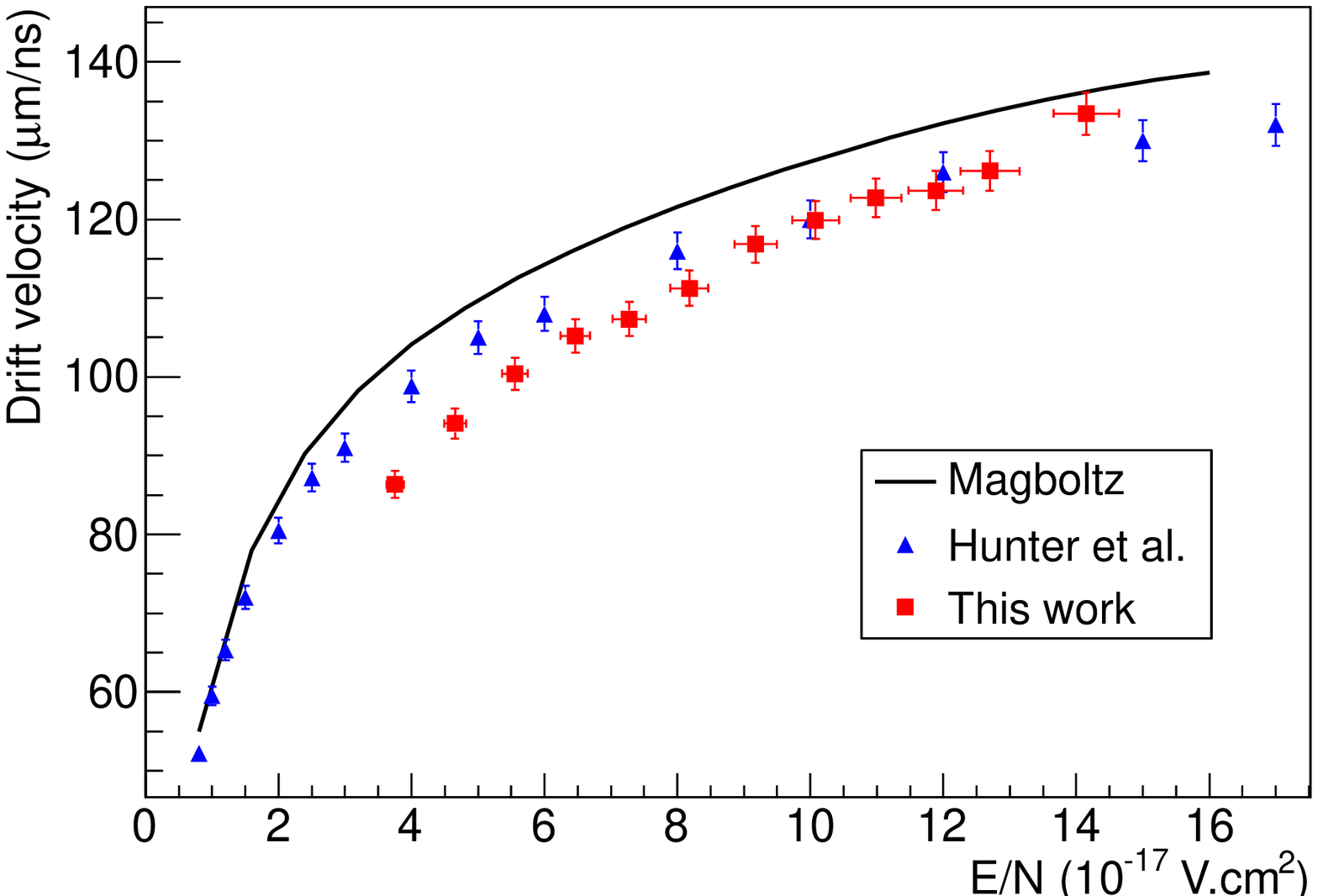}
\caption{\it Electron drift velocity measurement $v_d$ ($\mu {\rm m/ns}$) in a pure CF$_4$ gas at $50$~mbar as a  function 
of $E/N  \ (10^{-17} \ {\rm Vcm^{2}})$, for an
amplification field  $E_a = 14.5$ kV/cm. The data (red squares) have been obtained with the likelihood 
analysis strategy. We also present the Magboltz simulation (black line) and previous experimental data (blue triangle) \cite{hunter}.}
\label{fig:DriftVelocityCF4}
\end{center}
\end{figure}

Figure~\ref{fig:DriftVelocityCF4} presents the electron drift velocity measurement in a pure CF$_4$ gas at $50$~mbar, for an
amplification field  $E_a = 14.5$ kV/cm and a drift field $E_d$ ranging between 50 V/cm and 175 V/cm. 
The data (red squares) have been obtained with the likelihood analysis strategy presented in section \ref{sec:derive:vraisemblance}. 
The results obtained from a  Magboltz simulation \cite{magboltz} are also presented and compared with
previous experimental data \cite{hunter}. 
It can be first noticed that the measured electron drift velocity increases with increasing the drift field, from 
$v_d = 86.4$ $\mu$m/ns at   $E_d = 47$ V/cm to $v_d = 133.4$ $\mu$m/ns at $E_d = 175$ V/cm. There is a good agreement with the Magboltz
simulation, although the experimental results lie systematically below the results from the simulations, with  a shift ranging between  
18\% at $E_d = 47$ V/cm and 2\% at $E_d = 175$ V/cm. 
In fact, these discrepancies with the Magboltz simulation are expected as the drift velocity measured accounts for real, but unknown, 
experimental conditions (impurities, field inhomogeneities, ...) and corresponds to a long drift distance. This highlights the need to
measure this key parameter in the case of long drift
distances.
 
\begin{figure}[p]
\begin{center}
\includegraphics[scale=0.6,angle=0]{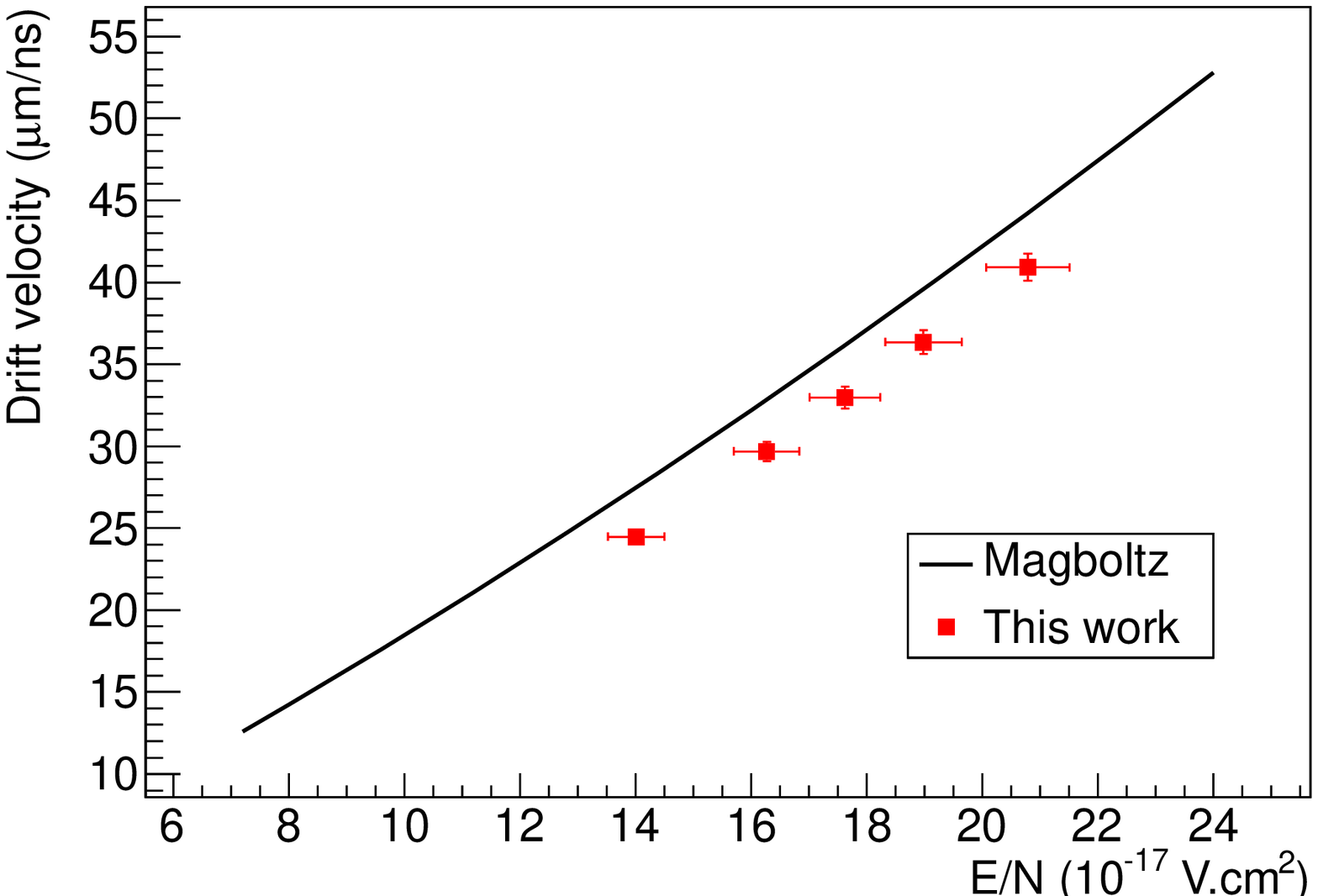} 
\caption{\it Electron drift velocity measurement $v_d$ ($\mu {\rm m/ns}$) 
in a 70\%CF$_4$ + 30\%CHF$_3$ gas mixture at $50$~mbar as a function of $E/N  \ (10^{-17}  \ {\rm Vcm^{2}})$, for an
amplification field  $E_a = 15.6$ kV/cm. The data (red squares) have been obtained with the likelihood analysis strategy. 
We also present the Magboltz simulation
(black line). To our knowledge there is no other experimental data with this gas mixture.}
\label{fig:DriftVelocityCF4CHF3}
\end{center}
\end{figure}

We have chosen to operate the MIMAC detector with a 70\% CF$_4$ + 30\% CHF$_3$ 
gas mixture. Magboltz simulations have shown that adding CHF$_3$ to pure CF$_4$ will lower the electron drift velocity.
This is a key point for directional Dark Matter as the track sampling along the Z axis ({\it
i.e.} along the drift field) will be improved when adding $\rm CHF_3$ while keeping almost the same 
Fluorine content of the gas mixture. Fluorine is indeed a golden target for spin-dependent Dark Matter search as the spin content is
dominated by the unpaired proton \cite{albornoz,canonni}. As a light nuclei, it is also considered by 
most projects of directional detection \cite{white}. A fraction of 30\% was found to be  
adequate for our experimental set-up (gain, track length, ...).\\

Figure~\ref{fig:DriftVelocityCF4CHF3} presents the electron drift velocity measurement in a 70\% CF$_4$ + 30\% CHF$_3$ 
gas mixture at $50$~mbar, for an
amplification field  $E_a = 15.6$ kV/cm. We focus on drift field values close to the MIMAC operating conditions.  
As for the pure $\rm CF_4$ gas, 
the measured electron drift velocity increases with increasing the drift field, with a systematic downward shift with respect to the 
 simulation results, ranging between 12\% and 8\%. To our knowledge there is no other experimental data with this gas mixture. 
 As expected, the electron drift velocity in a 70\% CF$_4$ + 30 \% CHF$_3$ gas mixture   is lower than in the pure 
$\rm CF_4$ case, by a factor $\sim$ 5. With this measurement, we show that a fraction of $\rm CHF_3$ in the gas mixture lowers
the electron drift velocity and consequently lengthens the track length in the time domain. It follows that 3D track reconstruction is improved
\cite{billard.track} as well as the e/recoil discrimination \cite{billard.discri}.\\

We have measured an effective electron drift velocity, in two gas mixtures,   corresponding to the real experimental conditions 
of the MIMAC directional Dark Matter 
detector. This is an {\it in situ} measurement, {\it i.e.} done with the Dark Matter detector itself, that can be performed for 
instance during calibration runs in order to check this key parameter.

\section{Conclusion}
We have presented a  new method for {\it in situ} electron drift velocity measurement, 
using an alpha source and a profile likelihood analysis based on the modeling of the signal induced on the grid. In particular, we have shown that such analysis allows us to avoid
bias due to {\it e.g.}  electron diffusion, ion collection time and electronic readout.
Hence, the effective electron drift velocity,  {\it i.e.} in the whole drift space, is measured corresponding to the real 
experimental conditions of the upcoming MIMAC directional detector. 
Following this study, we suggest to add $\rm CHF_3$ to the standard $\rm CF_4$ gas
used for directional detection as it allows us to lower the electron drift velocity 
while keeping almost the same Fluorine content of the gas mixture. 
In the case of the MIMAC detector, a fraction of 30\% was found to be an
adequate fraction as it allows to significantly enhance the 3D track reconstruction while
conserving sufficiently dense primary electron clouds in order to keep a high nuclear recoil track detection efficiency.\\
This result is of main interest for other CF$_4$ time projection chamber detectors dedicated to directional detection of Dark Matter such as DMTPC \cite{dmtpc} and NEWAGE \cite{newage} for both improving the timing resolution of tracks and the background rejection \cite{Lopez:2013ah}.

\section{Acknowledgments}
J. B. acknowledges support from NSF Award PHY-0847342.


\end{document}